\documentclass{webofc}

\usepackage[varg]{txfonts}   
\usepackage{hyperref}
\usepackage{url}
\usepackage{lineno, blindtext, indentfirst}
\usepackage{graphicx, float}
\usepackage{subcaption} 
\hypersetup{colorlinks=true,citecolor=blue,urlcolor=blue,linkcolor=blue}

\newcommand{ \pXi }{$p$$-$$\Xi^{-}$}

\newcommand{\pOmega} {$p$$-$$\Omega^{-}$}
\newcommand{\apaOmega} {$\bar{p}$$-$$\overline{\Omega}^{+}$}
\begin{document}
    
%
\title{Search for the Strange Dibaryons with Baryon Correlations in Isobar Collisions at STAR}
%
%

\author{\firstname{Kehao} \lastname{Zhang for the STAR Collaboration}\inst{1}\fnsep\thanks{\email{khzhang@mails.ccnu.edu.cn}} 
}

\institute{Institute of Particle Physics and Key Laboratory of Quark \& Lepton Physics (MOE), \\Central China Normal University, Wuhan, 430079, China. 
\
          }

\abstract{ Dibaryons provide insight into the strong interaction beyond conventional hadrons. Strange dibaryons, containing strange quarks, are especially valuable for probing hyperon-nucleon ($YN$) and hyperon-hyperon ($YY$) interactions. We report measurements of $p$–$\Xi^{-}$ and $p$–$\Omega^{-}$ correlation functions in Ru+Ru and Zr+Zr collisions at $\sqrt{s_\mathrm{NN}}$ = 200 GeV. The analysis, using the Lednicky-Lyuboshitz formalism, yields scattering parameters that offer key implications for the possible formation of strange dibaryon bound states, particularly $H$ ($S = -2$) and $N\Omega$ ($S = -3$).
}
\maketitle
\section{Introduction}
\label{intro}
\setlength{\parindent}{0em}
The study of exotic multi-quark states, especially strange dibaryons, remains a fascinating topic in hadron physics and Quantum Chromodynamics (QCD)~\cite{Chen:2024eaq}. The H-dibaryon—a deeply bound six-quark state—was first proposed by Jaffe in 1977~\cite{Jaffe:1976yi}. Another candidate, the nucleon–$\Omega$ (N$\Omega$) dibaryon, has also attracted significant interest~\cite{Goldman:1987ma}. Experimental searches through invariant-mass analyses are challenging due to short lifetimes and difficulty distinguishing them from scattering states. Two-particle correlation measurements in high-energy heavy-ion collisions provide a powerful indirect probe. The correlation function reflects final-state interactions and quantum statistics, sensitive to near-threshold bound or resonant states.

In these proceedings, we present measurements of proton–$\Xi^-$ ($p$–$\Xi^-$) and proton–$\Omega^-$ ($p$–$\Omega^-$) correlation functions in relativistic heavy-ion collisions, related to the H-dibaryon and N$\Omega$ dibaryon searches. Using the Lednický–Lyuboshitz (LL) formalism~\cite{LLmodel}, we extract scattering lengths and effective ranges, discussing their implications for possible strange dibaryon bound states with strangeness $S = -2$ and $S = -3$.

\section{Results}

The correlation function (CF) is constructed from the relative momentum $k^{*}$ in the pair rest frame (PRF). Experimentally, it is defined as $C(k^{*}) = \mathcal{N} A(k^{*}) / B(k^{*})$, where $A(k^{*})$ is the same-event distribution, $B(k^{*})$ is the mixed-event background, and the normalization factor $\mathcal{N}$ is chosen such that $C(k^{*}) = 1$ at large $k^{*}$. At low $k^{*}$, the CF is sensitive to final-state interactions and can be theoretically expressed as $C(k^{*}) = \int d^{3}r^{*} \, S(r^{*}) \, |\Psi(r^{*}, k^{*})|^{2}$, where $S(r^{*})$ is the source function and $\Psi(r^{*}, k^{*})$ is the relative wave function of the particle pair. The CF can be parameterized using the LL model~\cite{LLmodel}.

\subsection{\pXi{} Correlation Function}

\begin{figure}[h]
\centering
\begin{minipage}[c]{0.5\textwidth}
    \includegraphics[width=\textwidth]{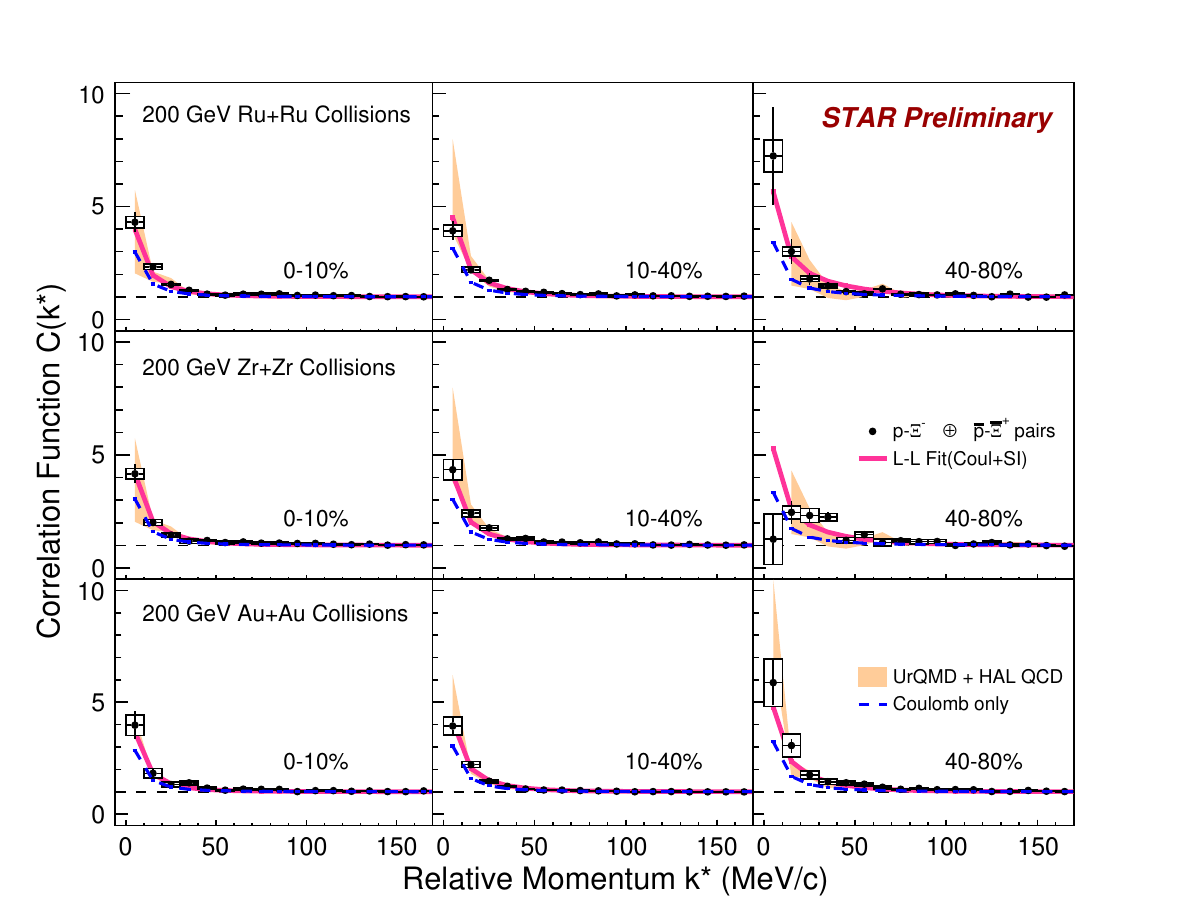}
\end{minipage}%
\hfill
\begin{minipage}[c]{0.5\textwidth}
    \caption{\pXi{} correlation functions in 0–10\%, 10–40\%, and 40–80\% centralities for Ru+Ru (top), Zr+Zr (middle), and Au+Au (bottom) at $\sqrt{s_{\mathrm{NN}}} = 200$ GeV. Black bars and boxes indicate statistical and systematic uncertainties, respectively. Magenta lines show simultaneous fits using the Lednický-Lyuboshitz model (including Coulomb and strong interactions), blue dashed lines show pure Coulomb, and orange band represents predictions obtained by combining HAL QCD potential with particle phase-space distributions generated by UrQMD~\cite{pxikamiya,pxiSasaki}.}
    \label{fig:pxiCF}
\end{minipage}
\end{figure}

\begin{figure}[htbp]
\vspace{-0.2cm}
\centering
\begin{minipage}[c]{0.5\textwidth}
    \includegraphics[width=\textwidth, clip]{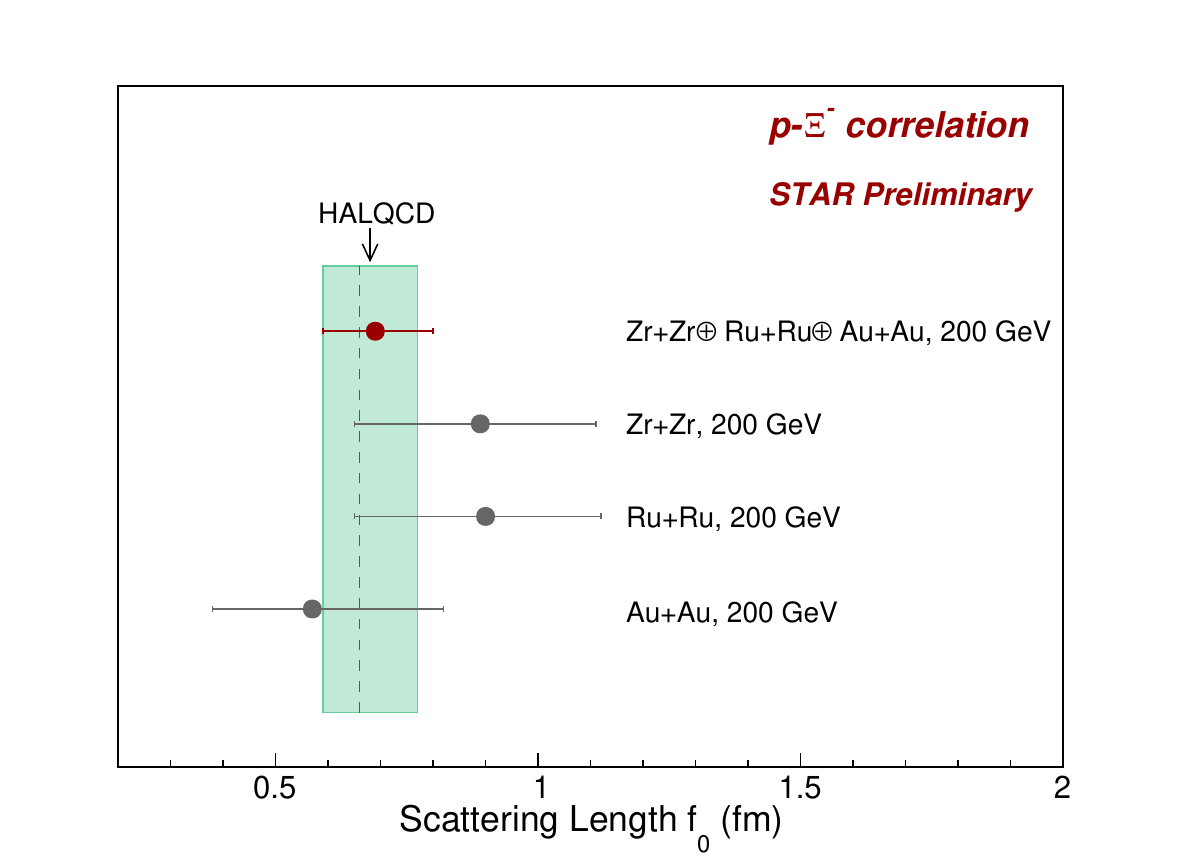}
\end{minipage}%
\hfill
\begin{minipage}[c]{0.45\textwidth}
    \caption{The spin-averaged \pXi{} scattering length parameter ($f_0$) extracted from simultaneous fits in Isobar and Au+Au collisions using the Bayesian method. The red point represent the results from simultaneous fits. The black points indicate the results for the three individual systems. The green band shows the prediction from HAL QCD~\cite{pxikamiya,pxiSasaki}.
    }
    \label{fig:contourpxi}
\end{minipage}
\vspace{-0.4cm}
\end{figure}

Figure~\ref{fig:pxiCF} shows the \pXi{} correlation functions as a function of $k^*$ in three centrality bins for Ru+Ru, Zr+Zr, and Au+Au collisions at $\sqrt{s_{\mathrm{NN}}} = 200$~GeV. An enhancement at low $k^*$ is observed in all centralities, becoming more pronounced in peripheral collisions, mainly due to the attractive Coulomb interaction. The data slightly exceed the Coulomb-only baseline, indicating a weakly attractive strong interaction. The LL model fits, incorporating both Coulomb and strong interactions, and HAL QCD predictions both describe the data well. Figure~\ref{fig:contourpxi} shows the scattering length ($f_0$) extracted from Bayesian simultaneous fits to the data in different centrality classes with a common set of parameters ($f_0$, $d_0$) using the LL framework. The positive $f_0$ supports a weakly attractive \pXi{} interaction and agrees with HAL QCD predictions~\cite{pxikamiya,pxiSasaki}.

\subsection{\pOmega{} Correlation Function}

\begin{figure}[h]
\vspace{-0.2cm}
\centering
\includegraphics[width=0.75\textwidth]{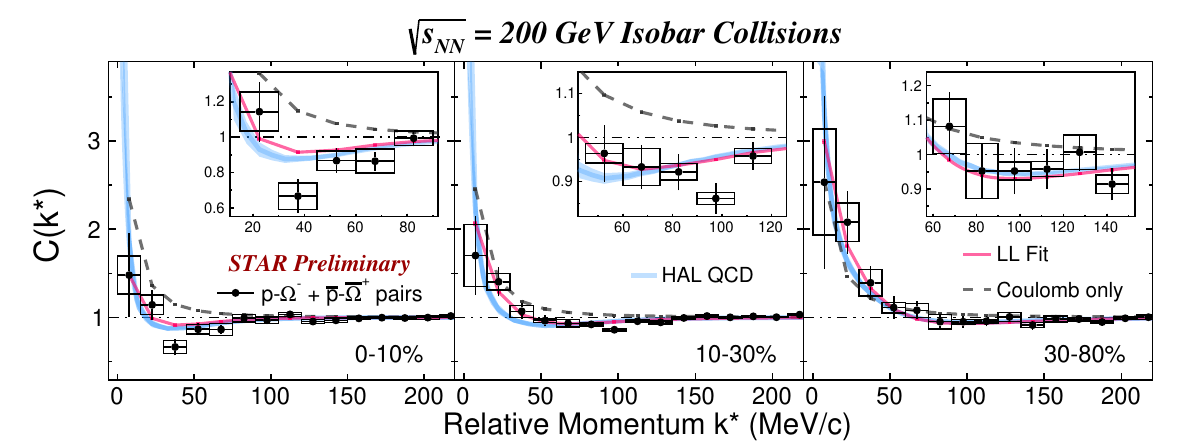}
\caption{Correlation functions for \pOmega{} and \apaOmega{} pairs measured in Ru+Ru and Zr+Zr collisions at $\sqrt{s_{NN}} = 200$ GeV. Open black symbols represent the data with statistical (bars) and systematic (boxes) uncertainties. The magenta lines show fits using the Lednický-Lyuboshitz model with the spin-averaged method, while gray dashed lines represent Coulomb-only contributions. Blue bands indicate HAL QCD predictions~\cite{HALQCD:2014okw,Morita:2019rph}. The insets are showing a zoom into the region near unity.}
\vspace{-0.4cm}
\label{fig:poCF}       
\end{figure}

\begin{figure}[htbp]
\vspace{-0.1cm}
\centering
\includegraphics[width=0.7\textwidth,clip]{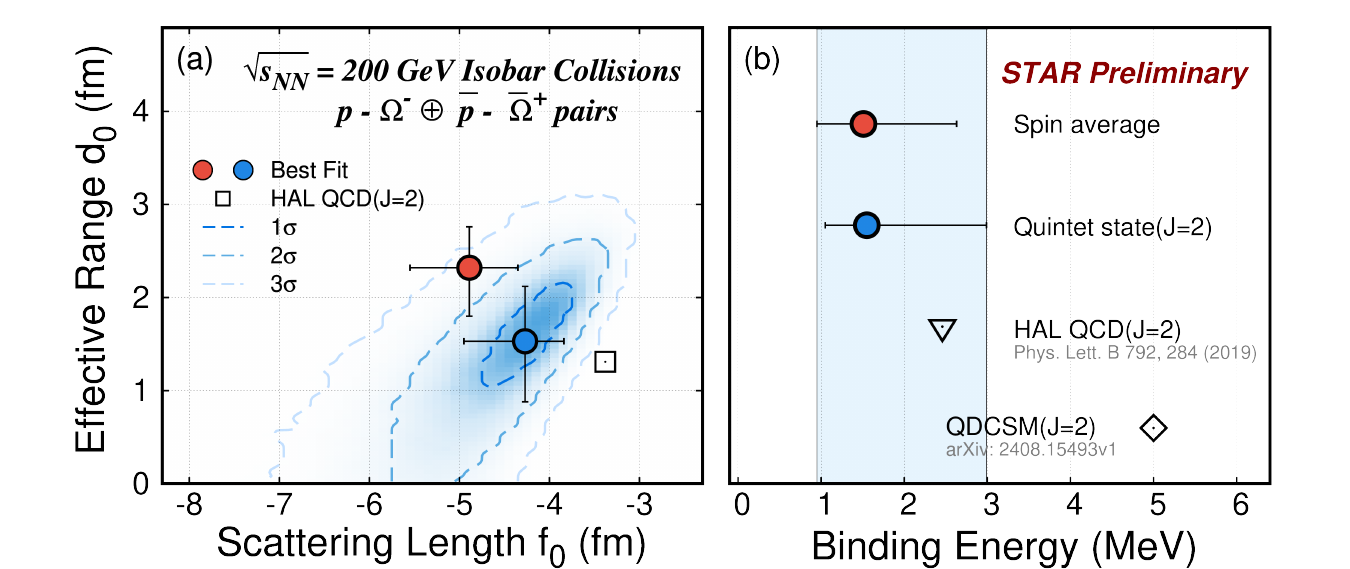}
\caption{(a) The extracted \pOmega{} scattering parameters, scattering length ($f_0$) and effective range ($d_0$), are shown as probability contours from spin-averaged (red) and quintet (blue) methods. Solid points mark best fits and blue bands show 1–3$\sigma$ confidence levels from the quintet method. (b) The \pOmega{} binding energy (BE), calculated via the Bethe formula, is shown with experimental points from spin-averaged (red) and quintet (blue) fits. HAL QCD prediction and QDCSM calculation are indicated by yellow inverted triangle and diamond markers, respectively~\cite{HALQCD:2018qyu,Yan:2024aap}.
}
\vspace{-0.5cm}
\label{fig:contourpo}     
\end{figure}

Figure~\ref{fig:poCF} shows the \pOmega{} correlation functions in Ru+Ru and Zr+Zr collisions at $\sqrt{s_{\mathrm{NN}}} = 200$~GeV. Similar to \pXi{}, an enhancement appears at low $k^*$ together with a suppression below unity, most evident in peripheral events (see insets). This suppression suggests strong-interaction effects from either a repulsive core or a bound-state node. The data are fitted with the LL model in a spin-averaged scheme including both Coulomb and strong interactions, and are well described by HAL QCD predictions. Figure~\ref{fig:contourpo} presents the extracted scattering length ($f_0$) and effective range ($d_0$) from both spin-averaged and quintet-channel fits—the latter treating quintet states with strong plus Coulomb interactions and triplet states with Coulomb only. The right panel shows the binding energy (E) from the Bethe formula, indicating a shallow bound state consistent with HAL QCD predictions~\cite{HALQCD:2014okw,Morita:2019rph}. The spin-averaged fit gives $f_0 = -4.9^{+0.5}_{-0.7}~\mathrm{fm}$, $d_0 = 2.3^{+0.4}_{-0.5}~\mathrm{fm}$, and $E = 1.5^{+1.1}_{-0.6}~\mathrm{MeV}$, while the quintet fit yields $f_0 = -4.3^{+0.4}_{-0.7}~\mathrm{fm}$, $d_0 = 1.5^{+0.7}_{-0.6}~\mathrm{fm}$, and $E = 1.6^{+1.4}_{-0.5}~\mathrm{MeV}$.

\subsection{Extracted Source and Scattering Parameters}

\begin{figure}[htbp]
\centering
\begin{minipage}[t]{0.48\textwidth}
    \centering
    \includegraphics[width=\linewidth, clip]{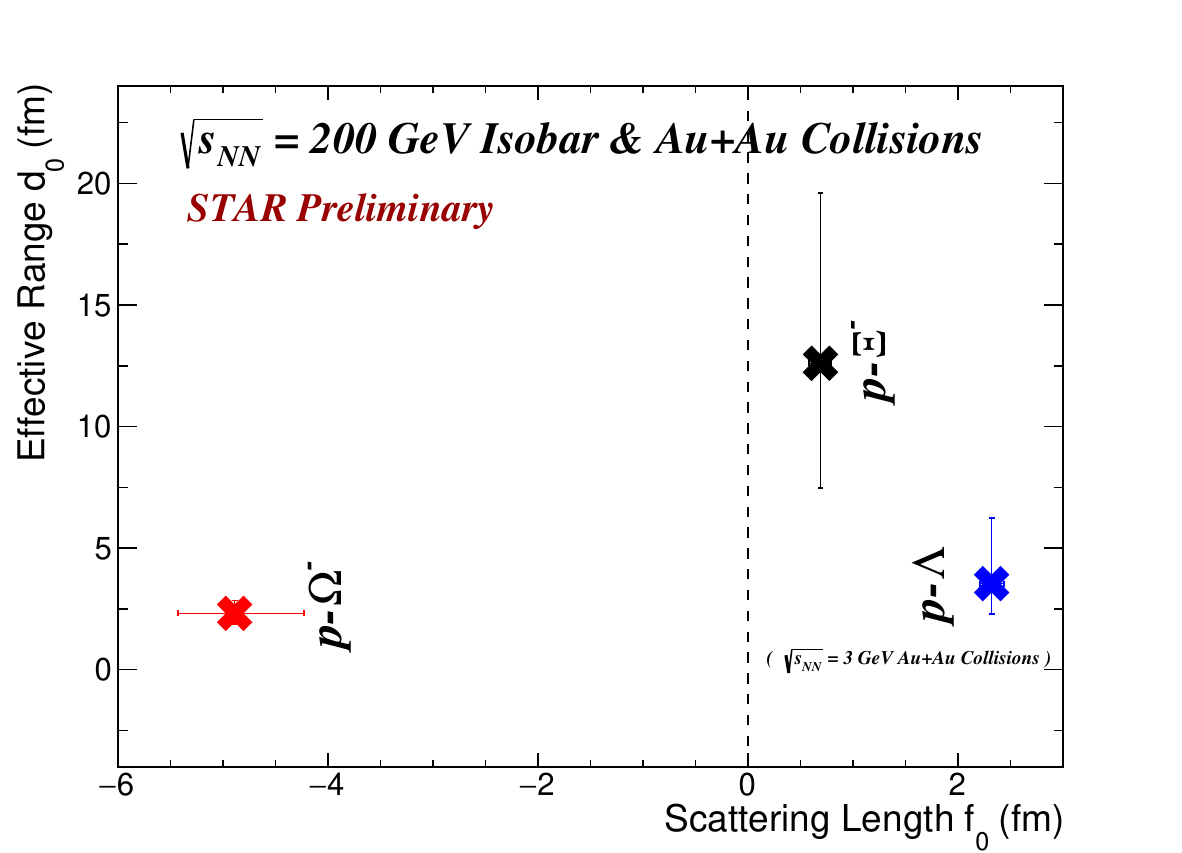}
    \caption{Extracted final state interaction parameters: scattering length ($f_0$) and effective range ($d_0$) for $p$--$\Lambda$ (blue), \pXi{} (black) and \pOmega{} (red) pairs.}
    \label{fig:summaryf0}
\end{minipage}
\hfill
\begin{minipage}[t]{0.48\textwidth}
    \centering
    \includegraphics[width=\linewidth, clip]{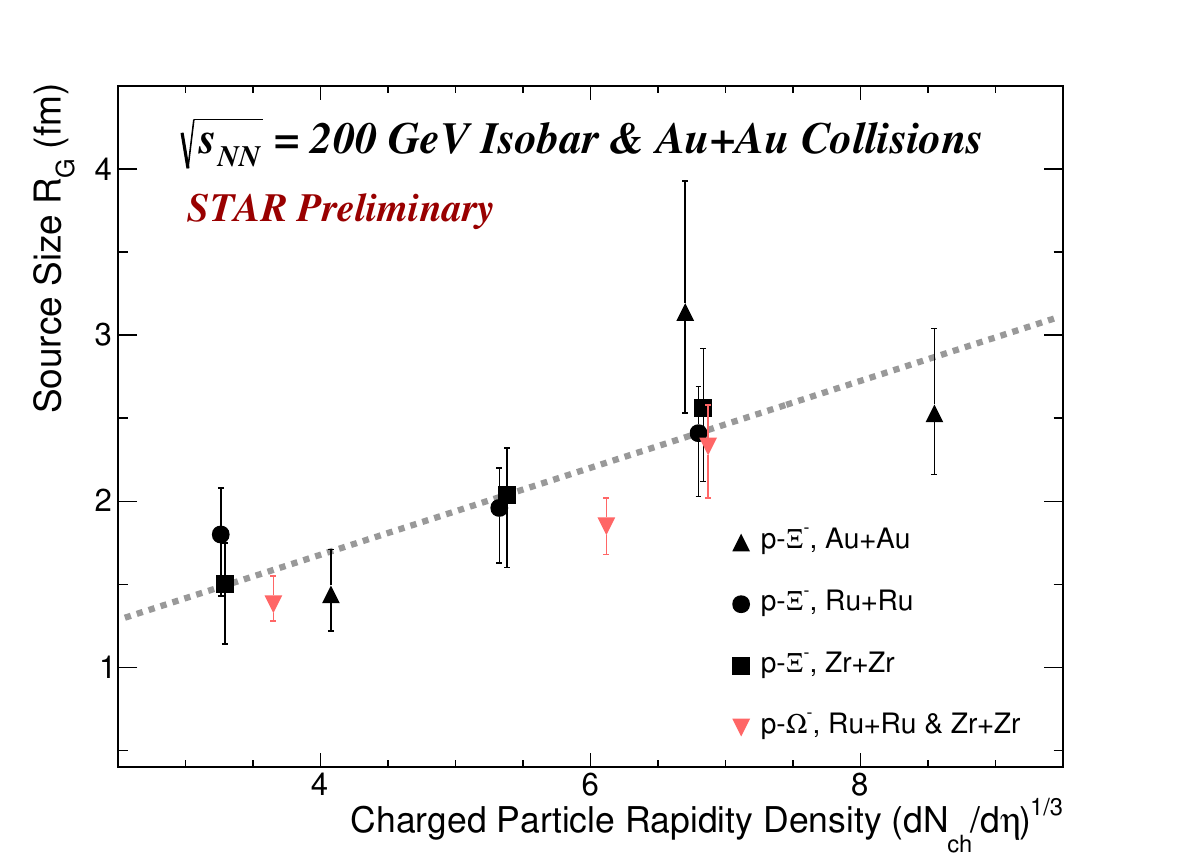}
    \caption{Extracted source size $R_{G}$ parameter as a function of $\left( \frac{dN_{\mathrm{ch}}}{d\eta} \right)^{1/3}$ for \pXi{} and \pOmega{} pairs in different collision systems.  }
    \label{fig:summaryr}
\end{minipage}
\end{figure}

Figure~\ref{fig:summaryf0} shows the extracted scattering parameters for the $p$--$\Lambda$, \pXi{}, and \pOmega{} pairs, obtained via a Bayesian fit using the LL model~\cite{Bayesian}. A positive scattering length ($f_0$) is observed for the $p$--$\Lambda$ and \pXi{} systems, indicating a weakly attractive interaction. In contrast, the $p$--$\Omega^-$ pair exhibits a negative scattering length, which supports the formation of a bound state. Figure~\ref{fig:summaryr} presents the extracted source sizes as a function of $\left( \frac{dN_{\mathrm{ch}}}{d\eta} \right)^{1/3}$, which correlates with collision centrality. The extracted source sizes fall within a reasonable range and show a clear centrality dependence, with more central collisions corresponding to larger source sizes.

\section{Summary}
\label{summary}
\setlength{\parindent}{0em}
In these proceedings, we present measurements of the \pXi{} and \pOmega{} correlation functions in Isobar collisions at $\sqrt{s_{\mathrm{NN}}} = 200$~GeV. A mild enhancement above the Coulomb baseline observed in the \pXi{} channel suggests a weakly attractive interaction, consistent with predictions from the HAL QCD model. Most notably, a significant suppression at low relative momentum ($k^*$) in the \pOmega{} channel provides the first experimental evidence for a shallow bound state. The extracted negative scattering lengths---$f_0 = -4.9^{+0.5}_{-0.7}$~fm (spin-averaged) and $f_0 = -4.3^{+0.4}_{-0.7}$~fm (quintet)---along with the derived binding energy $E=1.6_{-0.5}^{+1.4}$ MeV (3.2$\sigma$, spin state J=2), strongly support the formation of this strange dibaryon with high statistical significance. This represents the first experimental observation of a strange dibaryon.

\section{Acknowledgements}
This work was supported by National Key Research and Development Program of China (No.2022YFA1604900), National Natural Science Foundation of China (NO. 12525509 and 12447102).

%

\end{document}